\documentclass[aps,reprint,groupedaddress]{revtex4-1}

\usepackage{amsmath}
\usepackage{bm}
\usepackage{amssymb}
\usepackage{color}
\usepackage{multirow}
\usepackage[colorlinks,bookmarks=false,citecolor=red,linkcolor=blue,urlcolor=blue]{hyperref}
\usepackage{graphicx}
\usepackage{mathrsfs,bbm}
\usepackage{wrapfig}
\usepackage[version=3]{mhchem}
\usepackage{float}

\usepackage{todonotes}
\usepackage{empheq}
\usepackage[most]{tcolorbox}

\usepackage{tikz}   
\usepackage{tikz-3dplot} 

\newtcbox{\mymath}[1][]{
	nobeforeafter , math upper,tcbox raise base , enhanced , colframe=blue!30!black ,
	colback=blue!30,boxrule=1pt,#1 }

\usepackage{dcolumn}

\usepackage{calligra}
\DeclareMathAlphabet{\mathcalligra}{T1}{calligra}{m}{n}
\DeclareFontShape{T1}{calligra}{m}{n}{<->s*[2.2]callig15}{}

\def\Jhat{\hat{J}}
\def\Hhat{\hat{H}}

\def\Shat{\hat{S}}
\def\Uhat{\hat{U}}
\def\ahat{\hat{a}}
\def\bhat{\hat{b}}

\def\nhat{\hat{n}}

\begin{document}
\title{Power-law growth of time and strength of squeezing near quantum critical point}
	\author{Deepti Sharma} 
	\author{Brijesh Kumar}
	\email{bkumar@mail.jnu.ac.in}
	\affiliation{School of Physical Sciences, Jawaharlal Nehru University, New Delhi 110067, India.}
	\date{\today}  
		
\begin{abstract}
The dynamics of squeezing across quantum phase transition in two basic models, viz., the 
one-axis twisting model in transverse field and the Dicke model, is investigated using Holstein-Primakoff representation in the large spin limit. Near the phase boundary between the disordered (normal) and the ordered (superradiant) phase, the strength of spin and photon squeezing and the duration of time for which the system stays in the highly squeezed state are found to exhibit strong power-law growth with distance from the quantum critical point. The critical exponent for squeezing time is found to be 1/2 in both the models, and for squeezing strength, it is shown to be 1/2 in the one-axis twisting model, and 1 for the Dicke model which in the limit of extreme detuning also becomes 1/2.
\end{abstract}		

\maketitle

\section{Introduction}
The squeezed quantum states are of great practical importance owing to their usefulness in high precision measurements and many-particle entanglement~\cite{Wineland1992, Sorensen2001, Wang-Sanders, polzik2008quantum, Nori2011}. With its inception for photons~\cite{Loudon-Knight,Andersen2016}, the idea of squeezing to reduce quantum uncertainties (respecting Heisenberg's principle) has been suitably extended to atoms, or equivalently the spins~\cite{Wineland1992,Kitagawa1993,Wineland1994}, and is a vigorously pursued subject of research~\cite{Monika2010,exp-Leroux2010,Riedel2010,exp-Muessel2015,Zhang2015}. Among the many aspects of the studies carried out on spin squeezing~\cite{Law2001,Jin2007,Trail2010,Liu2011,Norris2012,Chaudhry2012,Zhang2015}, 
the one that interests us here is the possibility of its enhancement near a critical point~\cite{Vidal2004-QPT,Liu_2013,ABB2017,ABBhatta2017,Frerot2018,Balazadeh2018}. 

A model that has been widely used to investigate spin squeezing is the one-axis twisting (OAT) model~\cite{Kitagawa1993} in a transverse control field~\cite{Law2001}. More commonly, it is known as the infinite-range transverse field Ising model or the Lipkin-Meshkov-Glick (LMG) model~\cite{LMG}. It describes a system of $N$ spin-1/2's, where every spin interacts with every other spin via Ising interaction, and the external control field is perpendicular to the Ising direction. For the ferromagnetic interaction, in the large $N$ limit, it undergoes a continuous quantum phase transition from the disordered to ordered phase, driven by the competition between the control field and the interaction~\cite{Botet1982}. The OAT or LMG model can also be derived as an effective model from the Dicke model, which is a basic and widely studied model of quantum optics. The Dicke model describes a collection of $N$ two-level atoms interacting with a single mode of quantized radiation in a cavity~\cite{Dicke1954}. It exhibits a continuous phase transition from the normal (disordered) to the superradiant (ordered) phase~\cite{Hepp1973,Baumann2010,Kirton2019}. 

These two paradigmatic models are the objects of our study in this paper on spin squeezing across quantum phase transition. Here, we study the time evolution of spin squeezing in the OAT model in transverse field, and that of spin and photon in the Dicke model, to understand the pattern of growth in the strength and life of squeezing as one approaches the quantum critical point. We do this by employing Holstein-Primakoff representation~\cite{HP} for the spin operators to formulate solvable bosonic theories for the two models in the large spin limit. From this, we obtain the exact power-law by which the squeezing grows upon approaching the quantum phase boundary in the two models. We show that for a small (infinitesimal) parametric distance, $\delta$, from the quantum critical point, the duration of squeezing increases as $\delta^{-1/2}$ in both the models. The strength of squeezing is also shown to be governed by the same power law for the OAT model, and a similar power law (with a shuttle difference) for the Dicke model. Thus, we show that in the close neighbourhood of a critical point (or a line), not only the degree of squeezing is greatly enhanced, but also the duration of time over which it stays so is greatly enhanced. This is a finding of practical importance. 

In Sec.~\ref{sec:OAT} of this paper, we investigate the dynamics of spin squeezing across the quantum critical point in the OAT model in transverse field. Then, we investigate the squeezing for spin as well as photon across the superradiant transition in the Dicke model in Sec.~\ref{sec:Dicke}. Notably, the photon squeezing in the Dicke model exhibits the same critical behaviour as the spin squeezing, but only in the oppositely detuned limits (i.e., not together). We conclude the paper with a summary in Sec.~\ref{sec:sum}.

\section{One-Axis Twisting Model in Transverse Field \label{sec:OAT}}
The Hamiltonian of the OAT model in the presence of transverse control field read as:
\begin{equation} \label{eq:OAT}
\hat H = -\kappa\hat{S}_{x}^2 + \Omega{\hat{S}_{z}} 
\end{equation}	 
where $\kappa$ is the strength of the one-axis twisting (or the Ising interaction) and $\Omega$ is the transverse field, both of which are taken to be positive in this paper. The operators $\Shat_x$ and $\Shat_z$ are the $x$ and $z$ components of a spin with quantum number $J$. We take $J$ to be large. 

It is known that this model for $\kappa>0$ undergoes a continuous phase transition in the ground state by changing $\Omega$ with respect to $\kappa$~\cite{Botet1982}. For large $\Omega$, the average spin will point along $-z$ direction in the ground state, whereas for large enough $\kappa$, it will also have a component along $\pm x$ direction. This quantum phase transition can be easily described by a mean-field theory as follows.

The order parameter, $m$, for this phase transition is given by the expectation value of $\Shat_x$ in the ground. That is, $m=\langle\hat{S}_x\rangle$. Under mean-field approximation, Eq.~\eqref{eq:OAT} becomes: $ \hat{H} \approx \kappa m^2 - 2\kappa m \hat{S}_x  + \Omega \hat{S}_z$. It can be diagonalized by the following spin-rotation for $\theta=\tan^{-1}\left(-\frac{2\kappa m}{\Omega}\right)$. 
\begin{subequations}
	\label{eq:rotationsfromStoJ}
	\begin{eqnarray}
	\hat{S}_{z}&= &\hat{J}_{z} \cos\theta - \hat{J}_{x} \sin\theta  \\
	\hat{S}_{x}&= &\hat{J}_{z} \sin\theta + \hat{J}_{x} \cos\theta 
	\end{eqnarray}	
\end{subequations}
Here, $\hat{J}_x$ and $\Jhat_z$ are the spin operators in the rotated frame. The ground state of this mean-field model is given by the eigenstate $|J,-J\rangle$ of operator $\Jhat_z$ with eigenvalue $-J$.
We obtain the order parameter, $m$, as given below, by calculating the expectation value of $\Shat_x$ in this mean-field ground state. 
\begin{equation}
m = \Bigg\{
\begin{array}{l l} 
\pm J\sqrt{1 -\xi^2}  &~\forall~~ \xi < 1 \\
0 & ~\forall~~ \xi \ge 1  
\end{array} 
\label{eq:OP-m}
\end{equation}
Here, $\xi= \frac{\Omega}{2\kappa J}$ is a dimensionless parameter. The rotation angle $\theta$, which measures the direction of the average spin vector in the two phases, is given below.
\begin{equation}
\theta = \Bigg\{
\begin{array}{l l} 
\cos^{-1}{\xi}  &~\forall~~ \xi < 1 \\
0 & ~\forall~~ \xi \ge 1  
\end{array} 
\label{eq:theta}
\end{equation}
Thus, for $\Omega < 2\kappa J$, the OAT model in transverse field realizes an ordered phase with two degenerate values for $m$. For $\Omega  \ge 2\kappa J$, it is in the disordered phase with $m=0$. It is clearly a continuous (second-order) quantum phase transition, because the order parameter changes continuously with $\xi$ across the critical point $\xi_c=1$. 

\subsection{Bosonic theory in the large spin limit \label{sec:OAT-HP}} 
We now develop a theory of quantum fluctuations in the two phases (ordered and disordered) discussed above. This is best done by using the Holstein-Primakoff (HP) representation for the spin operators~\cite{HP}, which affords a nice and solvable bosonic theory of quantum fluctuations in the large $J$ limit. Since the mean-field ground state in both the phases is given by state $|J,-J\rangle$ of the spin operator $\Jhat_z$, we write the following HP representation:
\begin{subequations}
\begin{eqnarray}
\hat{J}_{z} &=& -J +\hat{n} \label{eq:HP-Jz} \\
\hat{J}^+ &=& \hat{a}^\dagger \, \sqrt{2J-\hat{n}} \label{eq:HP-Jp}
\end{eqnarray}
\end{subequations}
for the spin operators in terms of the boson creation and annihilation operators $\ahat^\dag$ and $\ahat$. Here, $\nhat = \ahat^\dag\ahat$ is the boson number operator, and $\Jhat^- = (\Jhat^+)^\dag$. This representation describes the fluctuations with respect to the reference state $|J,-J\rangle$ as bosons with a constraint, $\nhat \le 2J$. In the large $J$ limit, Eq.~\eqref{eq:HP-Jp} approximates to $\Jhat^+ =\ahat^\dag \sqrt{2J}$, which greatly simplifies the representation. 

By rewriting Eq.~\eqref{eq:OAT} in the rotated frame using Eqs.~\eqref{eq:rotationsfromStoJ} for $\cos{\theta}=\xi$, and then applying the HP transformation in the large $J$ limit, we obtain the following bosonic Hamiltonian for the OAT model in transverse field. 
\begin{equation}
\hat H =   c_1 \hat{a}^\dagger\hat{a}+ c_2 \left(\hat{a}^\dagger\hat{a}^\dagger+\hat{a}\hat{a}\right) + c_3  
\label{eq:bosonicHamilt}
\end{equation}   
where the coefficients $c_1$, $c_2$, and $c_3$ are listed below. 
\begin{subequations}
\begin{eqnarray}
c_1 &=&  \Omega \cos{\theta} + 2\kappa J \sin^2{\theta} -\kappa J \cos^2{\theta} \\
c_2 &=& -\frac{\kappa J}{2} \cos^2{\theta} \\
c_3 &=& -J\left[\Omega \cos{\theta} +\kappa J\sin^2{\theta} + \frac{\kappa}{2}\cos^2{\theta} \right] 
\end{eqnarray}
\label{eq:c123}
\end{subequations}
Here, the angle $\theta$ in the two phases is given by Eq.~\eqref{eq:theta}.

The boson Hamiltonian of Eq.~(\ref{eq:bosonicHamilt}) can be diagonalized by the Bogolioubov transformation, $\hat{U}_a $, given below. 
\begin{equation}
\hat{U}^{ }_a = \exp\left[-\theta_a\left(\hat{a}^\dagger\hat{a}^\dagger -\hat{a}\hat{a} \right)\right]
\label{eq:Ua}
\end{equation}
Here, $\tanh{4\theta_a} = 2c_2/c_1$, for $c_1$ and $c_2$ given by Eqs.~\eqref{eq:c123}. 
The resulting diagonal Hamiltonian, $\mathscr{H}$, reads as:
\begin{equation} 
\hat{U}_a^\dagger \, \Hhat \, \hat{U}^{ }_a = \mathscr{H} =  \omega_a\left(\nhat+\frac{1}{2}\right)+c_3-\frac{c_1}{2}
\label{eq:diagonalOATboson}
\end{equation}
where 
\begin{align}
\omega_a & =   2\kappa J\sqrt{\big( \xi \cos\theta - \cos{2\theta} \big)\big( \xi \cos\theta + \sin^2{\theta}\big)} \nonumber \\ 
& =  2\kappa J \left\{
\begin{array}{l l} 
\sqrt{1 -\xi^2}  & ~~\forall~~ \xi < 1 \\
\sqrt{\xi\left(\xi-1\right)} & ~~\forall~~ \xi \ge 1  
\end{array} 
\right. \label{eq:omegaa}
\end{align}
is the energy of elementary bosonic excitations. Note that $\omega_a > 0$ inside the two phases, but it continuously tends to $0$ upon approaching the quantum critical point from either side. 
	
\subsection{Squeezing dynamics across phase transition}\label{sec:squeezingOAT-2}
In both the phases of the OAT model in transverse field, the average spin vector points along $\Jhat_z$. Therefore, as per the prescription enunciated by Kitagawa and Ueda~\cite{Kitagawa1993}, the minimum uncertainty in the spin component, $\hat{J}_\phi = \hat{J}_x\cos{\phi} + \hat{J}_y\sin{\phi}$, perpendicular to $\Jhat_z$ would determine the spin squeezing in the two phases.  For the uncertainty, $\Delta \hat{J}_\phi(t) = \sqrt{\langle \hat{J}^2_\phi (t) \rangle - \langle \hat{J}_\phi (t) \rangle^2 }$, in the value of $\Jhat_\phi$ in a quantum state at time $t$, the parameter
\begin{equation}\label{eq:squeezingForm}
\zeta_s = \frac{\Delta \hat{J}_\phi(t)_{\rm min}}{\sqrt{J/2}}
\end{equation}
quantifies  the spin squeezing as a function of time. Here, the subscript `min' stands for the minimum value of $\Delta\Jhat_\phi$ obtained by minimizing it with respect to $\phi$. A quantum state is said to be spin-squeezed only if $\zeta_s <1$. Closer the value of $\zeta_s$ to $0$, higher is the spin squeezing. Thus, the strength of spin squeezing can be quantified by $\zeta_s^{-1}$.

We investigate spin squeezing by calculating uncertainties in the time dependent state, $e^{-i\Hhat t} |0\rangle$, where $|0\rangle$ is the vacuum of the Holstein-Primakoff boson defined by Eq.~\eqref{eq:HP-Jz}, and the $\Hhat$ is given by Eq.~\eqref{eq:bosonicHamilt}. The vacuum state $|0\rangle$ is same as the mean-field spin ground state, $|J,-J\rangle$, in the two phases, and the $\Hhat$ describes the dynamics of quantum fluctuations of the mean-field state in the large $J$ limit.  

The operator $\Jhat_\phi$ introduced above can, in the large $J$ limit, be written as: $\hat J_\phi = \sqrt{\frac{J}{2}}\left(\ahat^\dag e^{-i\phi} + \ahat e^{i\phi}\right)$. One can check that $\langle \hat J_\phi(t)\rangle = \langle 0|e^{i\Hhat t}\, \hat J_\phi\,e^{-i\Hhat t}|0\rangle = 0$. Therefore, the spin squeezing parameter can be written as:
\begin{eqnarray}
\zeta_s(t) &=& \sqrt{\left\langle 0\right| e^{i\Hhat t} \left(\ahat^\dag e^{-i\phi} + \ahat\,e^{i\phi}\right)^2 e^{-i\Hhat t}\left|0\right\rangle}_{\rm min} \nonumber \\
&=& \sqrt{\left[ A_s(t)\cos{2\phi} + B_s(t)\sin{2\phi} + C_s(t)\right]}_{\rm min} \nonumber \\
&=& \sqrt{C_s(t)-\sqrt{A^2_s(t)+B^2_s(t)}} \label{eq:zetas-HP}
\end{eqnarray}
with minimization angle, $\phi_{\rm min}(t) = \frac{\pi}{2} + \frac{1}{2 } \tan^{-1}\left[\frac{B_s(t)}{A_s(t)} \right] $. Parameters $A_s(t)$, $B_s(t)$ and $C_s(t)$ are given below. 
\begin{subequations}
\label{eq:ABC}
\begin{eqnarray}
A_s(t)  &=& \langle 0|e^{i\Hhat t} \left(\ahat^\dag\ahat^\dag+\ahat\ahat\right)e^{-i\Hhat t}|0\rangle \\
&=& -\sin^2(\omega_at) \sinh(8\theta_a) \nonumber \\
B_s(t) & = & -i \langle 0|e^{i\Hhat t} \left(\ahat^\dag\ahat^\dag-\ahat\ahat\right)e^{-i\Hhat t}|0\rangle \\ 
&=& \sin(2{\omega_a} t) \sinh(4\theta_a)  \nonumber \\
C_s(t)  &=& 1+ 2 \langle 0|e^{i\Hhat t} \ahat^\dag\ahat e^{-i\Hhat t}|0\rangle \\
& = & \cos^2(\omega_at) + \sin^2(\omega_a t)\cosh(8\theta_a) \nonumber
\end{eqnarray}
\end{subequations}
With this, we are set to calculate and discuss spin squeezing in the ordered and disordered phases. 

\subsubsection{Ordered phase}
In the ordered phase for $\xi<1$, we can write $\xi=1-\delta$, where $0<\delta<1$. Note that $\delta=0$ corresponds to the critical point. Hence, $\delta$ can act as a small expansion parameter close to the critical point. The frequency $\omega_a$ [see Eq.~\eqref{eq:omegaa}] and the Bogoliubov angle $\theta_a$ [see Eq.~\eqref{eq:Ua}] in the ordered phase can be written as follows.  
\begin{subequations}
\begin{eqnarray}
\omega_{a} & =&  2\kappa J\sqrt{\delta(2-\delta)} \label{eq:omegaaO}\\
\theta_a & =  & \frac{1}{8}\left[\log{\delta} + \log(2-\delta)\right] \label{eq:thetaaO}
\end{eqnarray}
\end{subequations}		
We put this $\omega_a$ and $\theta_a$ in Eqs.~\eqref{eq:ABC} for $A_s$, $B_s$ and $C_s$ to calculate the time dependent spin squeezing parameter, $\zeta_s$, as defined in Eq.~\eqref{eq:zetas-HP}. In our calculations, we measure frequency (energy) in units of $2\kappa J$, and therefore time in units of $1/2\kappa J$.
	
\begin{figure}[h!]
		\centering
		\includegraphics[width=.5\textwidth]{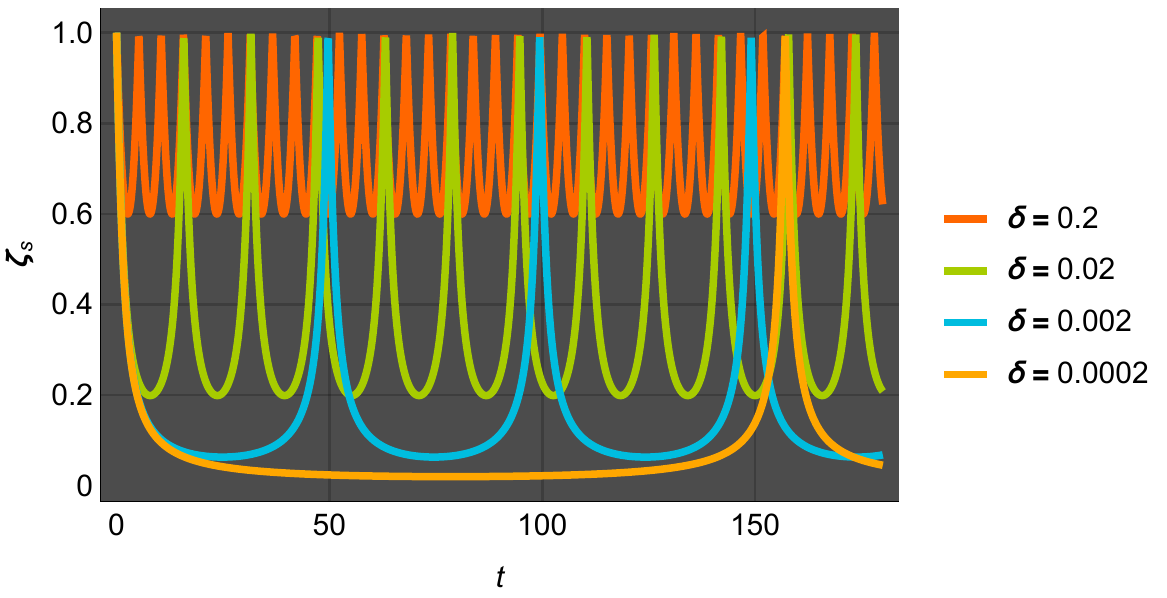}
		\caption{Spin squeezing, $\zeta_s$ vs. time, in the ordered phase of the OAT model in transverse field for $\delta=0.2, 0.02, 0.002$ and $0.0002$. As $\delta$ gets closer to $0$ (the quantum critical point), the minimum value of $\zeta_s$ also gets smaller. Very close to the critical point, $\zeta_s$ stays nearly flat for most part of every cycle,  at a very small value (i.e., strong squeezing with a long life).}
		\label{fig:OrdSqzpart2}	
	\end{figure}

The $\zeta_s$ vs. $t$ for different values of $\delta$ are presented in Fig.~\ref{fig:OrdSqzpart2}. It exhibits an oscillatory behaviour with time, because the time evolution here is governed by purely unitary dynamics (as we have neglected dissipation to keep it simple). Upon decreasing $\delta$, we see a systematic reduction in the minimum value of $\zeta_s$ (which is always smaller than 1), and also a systematic increase in the time period, $T$, of oscillation. Remember that a smaller value of $\zeta_s$ implies more squeezing. Moreover,  a longer $T$ would imply a longer life for the system to stay in a squeezed state. A prominent feature of this data is that, very close to the quantum critical point ($\delta \rightarrow 0$), the greatly reduced $\zeta_s$ tends to become nearly flat as a function of $t$ with a greatly elongated time period. What it tells is that the system is able to realize and sustain a highly squeezed state over a great length of time! This is indeed a finding of some practical value.

The longer life of squeezing near the critical point is easier to understand. The explicit time dependence in Eqs.~\eqref{eq:ABC} implies that the time period of oscillation of squeezing is given by $\omega_a T=\pi$. From Eq.~\eqref{eq:omegaaO} for $\omega_a$, it is clear that $T = \pi/\sqrt{\delta(2-\delta)}$ for any $\delta$, which for $\delta\rightarrow 0$ implies that $T\approx \pi/\sqrt{2\delta}$. Thus, very close to the critical point, the squeezing time grows as $\delta^{-1/2}$.				

The reduction in the value of $\zeta_s$ with $\delta$ can be understood by noting that the minimum of $\zeta_s$ in every squeezing cycle occurs at the half-time period, $T/2$. We find that $\zeta_s(T/2) = \sqrt{\delta(2-\delta)}$ for any $\delta$. Thus, very close to the critical point, the minimum of $\zeta_s$ reduces as $\delta^{1/2}$. To understand the emergence of the flatness of $\zeta_s$ in time (for most part in every squeezing cycle) close to the critical point, we expand $\zeta_s(t)$ around $T/2$. Let the time in a squeezing cycle be redefined as $t=T/2+ \tau$, where $|\tau| < \frac{T}{2} $. For a small $\tau$, and a small $\delta$, we obtain the following series expansion that applies to the $\zeta_s$ in the middle portion of a squeezing cycle.  
\begin{equation}\label{eq:ExpansDeltaOrdered} 
\zeta_s\left(\frac{T}{2}+\tau\right) = \sqrt{2\delta}\left[1 + \delta \tau^2 +\frac{5}{6}\delta^2 \tau^4 + \frac{61}{90}\delta^3 \tau^6+
\dots\right]
\end{equation}	
Notably, in Eq.~\eqref{eq:ExpansDeltaOrdered}, the terms with higher powers of $\tau$ tend to become insignificant faster as $\delta$ tends to zero. It means that, in the close proximity of the quantum critical point, the leading behaviour of squeezing is dominated by the first term in the power series, i.e., $\zeta_s \approx \sqrt{2\delta}$, which is independent of time. This explains the flattening of the squeezing curve in Fig.~\ref{fig:OrdSqzpart2} for very small values of $\delta$.

\subsubsection{ Disordered phase }
\begin{figure}[h!]
	\centering
	\includegraphics[width=.5\textwidth]{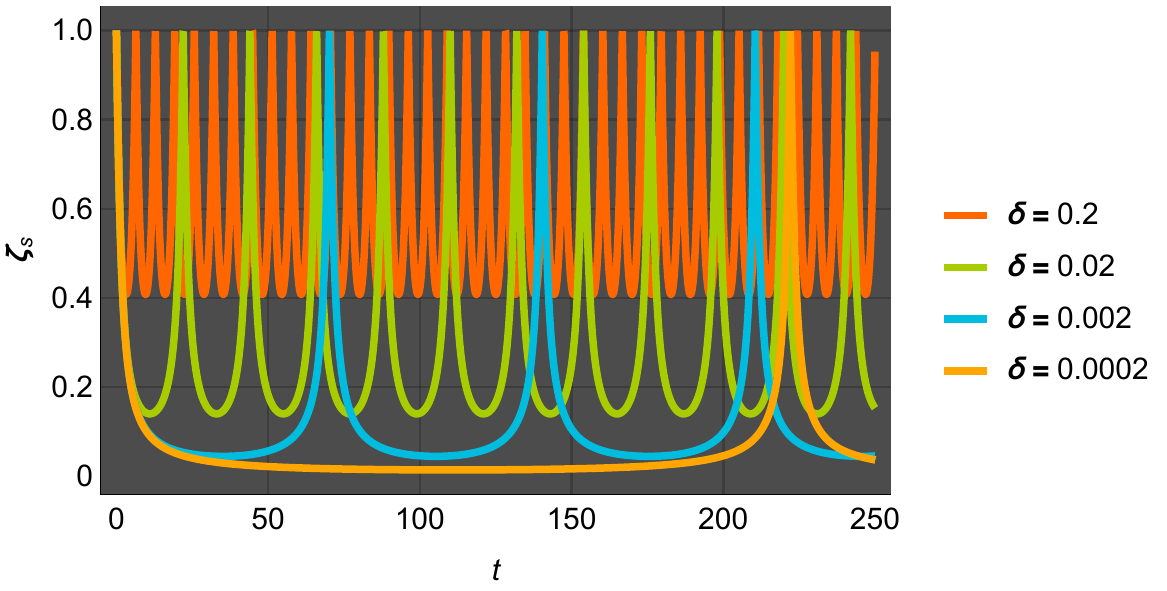}
	\caption{Spin squeezing, $\zeta_s$ vs. time, in the disordered phase of the OAT model in transverse field for $\delta=0.2, 0.02, 0.002$ and $0.0002$. Notice the prominent reduction in $\zeta_s$, and its nearly flat behaviour, very close to quantum critical point.}
	\label{fig:disordSqzpart2}		
\end{figure}

In the disordered phase, we can write $\xi=1+\delta$, where $\delta>0$. The $\delta$ dependence of frequency and Bogoliubov angle in this phase is given below. 
\begin{subequations}
\begin{eqnarray}
\omega_a &=& 2\kappa J \sqrt{\delta(1+\delta)} \label{eq:omegaaD}\\
\theta_a &=& \frac{1}{8}\left[\log{\delta}-\log(1+\delta)\right] \label{eq:thetaaD}
\end{eqnarray}	
\end{subequations}
By putting these into Eqs.~\eqref{eq:ABC}, we obtain $\zeta_s$ vs. $t$ for different values of $\delta$, as plotted in Fig.~\ref{fig:disordSqzpart2} (with $t$ in units of $1/2\kappa J$).	
As in the ordered phase, here too, we see a strong enhancement in the life and strength of squeezing upon approaching the quantum critical point.
Similar to Eq.~\eqref{eq:ExpansDeltaOrdered}, the following expansion describes the flattened behaviour of squeezing in the middle portion of every squeezing cycle for small $\delta$ and small $\tau$ (i.e., time measured from the half-time period, $T/2$).
\begin{equation}\label{eq:ExpansDeltaDisordered}  
 \zeta_s\left(\frac{T}{2}+\tau\right) = \sqrt{\delta}\left[1 + \frac{1}{2}\delta \tau^2 +\frac{5}{24} \delta^2 \tau^4 + \frac{61}{720} \delta^3 \epsilon^6 + \cdots \right]
\end{equation}
Here, the time period of squeezing oscillation is given by $T=\pi/\sqrt{\delta(1+\delta)}$. 

To summarise the important finding of this section, the enhancement of spin squeezing near quantum critical point ($\xi_c=1$) is described by the following power laws. 
\begin{subequations}
\label{eq:powerlawOAT}
\begin{eqnarray}
\mbox{Squeezing time ($\propto T$)} & \sim & |\xi-\xi_c|^{-1/2} \label{eq:T-powerlawOAT} \\
\mbox{Minimum value of $\zeta_s$} & \sim & |\xi-\xi_c|^{1/2} \label{eq:zetasmin-powerlawOAT}
\end{eqnarray} 
\end{subequations}
Although this result is obtained for the OAT model in transverse field, i.e. Eq.~\eqref{eq:OAT}, but the power-law growth of squeezing time and strength in the close proximity of a quantum critical point can be a more general possibility. 

\section{Dicke Model}\label{sec:Dicke}
Motivated by the findings in the previous section, we now investigate in Dicke model the interesting possibility of critically enhanced squeezing lasting for longer times across the quantum phase transition from the normal to superradiant phase. The Dicke model written below describes the physics of $N$ two-level atoms interacting with a single mode of quantized radiation inside a cavity~\cite{Dicke1954}. 
\begin{equation}
\hat H = \omega\hat{a}^\dagger\hat{a} + \epsilon{\hat{S}_{z}}+\frac{g}{\sqrt{N}}(\hat{a}^\dagger+\hat{a}){\hat{S}_{x}} 
\label{eq:Dicke}
\end{equation}
Here, $\omega$ is the frequency of a radiation mode in the cavity, and $\epsilon$ is the transition frequency of the two-level atoms. The collective atomic variables are the spin operators $\Shat_z$ and $\hat{S}_{x}$, with spin quantum number $J=N/2$, of which the $\Shat_x$ describes the atomic dipole and $\hat{S}_{z}$ measures the population difference of atoms in the excited and ground state. The dipole interaction between the radiation and the atoms is denoted by $g$. The operators $\hat{a}$ ($\hat{a}^\dagger$) are the annihilation (creation) operators of the quantized radiation (photons).  Some generalized versions of the Dicke model are also current~\cite{Dimer2007,Nagy2010,Bhaseen2012}, but here we keep it simple by working on its basic form as in Eq.~\eqref{eq:Dicke}. 

\subsection{Superradiant transition in the ground state \label{sec:Dicke-MFT}}
The Dicke model is known to exhibit a quantum phase transition from the normal phase (with atoms in their ground state) to the superradiant phase (wherein a mascroscopic fraction of atoms are kept in their excited state cooperatively by the radiation)~\cite{Hepp1973}. The atoms and photons in the superradiant phase spontaneously develop two order parameters: $\langle\hat{S}_x\rangle = m \neq0 $ and $\langle\hat{a}^\dagger\rangle=\langle\hat{a}\rangle = \alpha \neq0 $. 
One can determine $m$ and $\alpha$ by doing the mean-field theory outlined below.

\begin{figure}[h!]
	\centering
	\includegraphics[width=.4\textwidth]{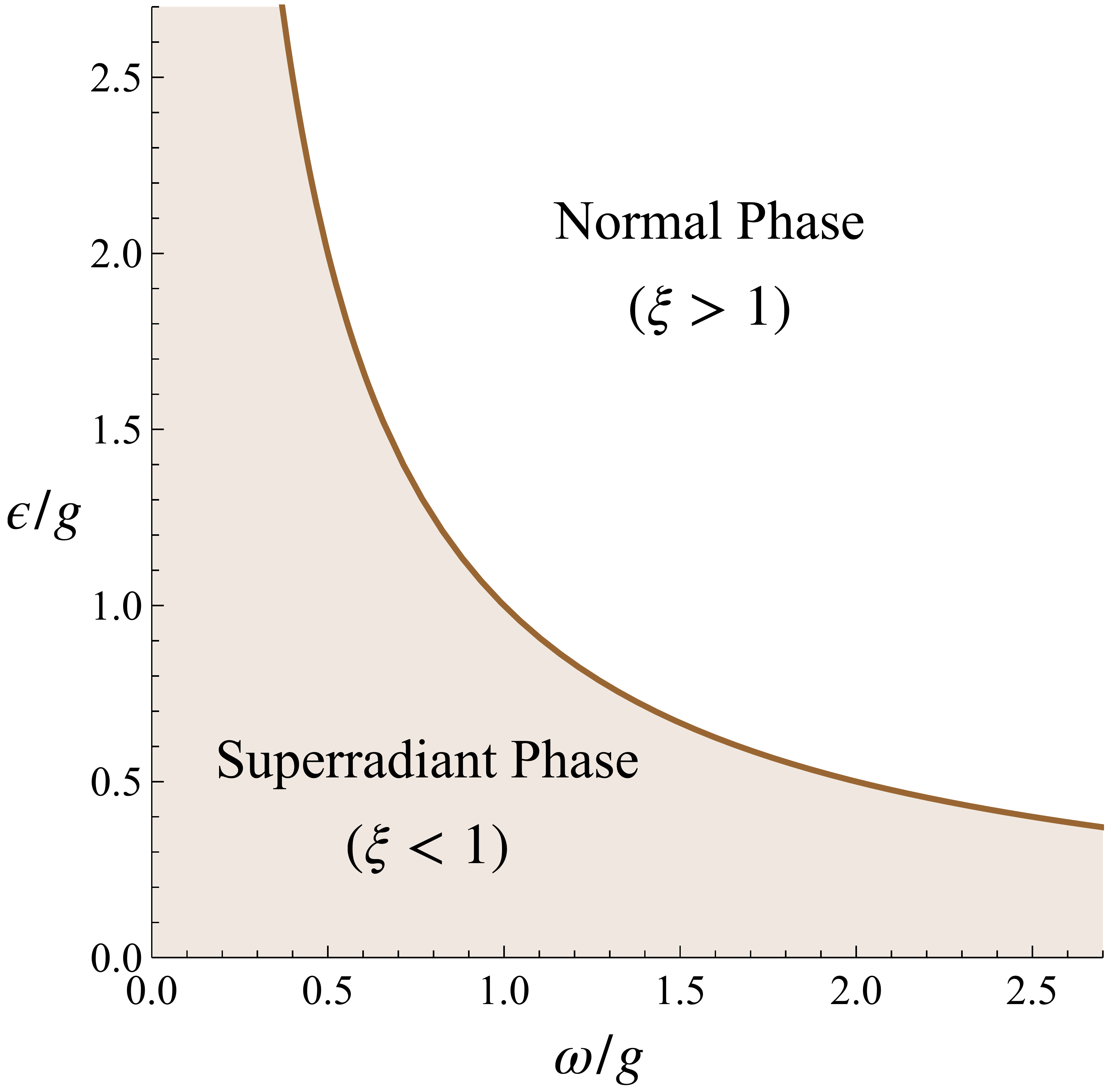}
	\caption{Quantum phase diagram of the Dicke model in mean-field theory. The boundary between the superradiant and normal phase is given by the critical line $\frac{\epsilon \, \omega}{g^2}=1$.}
	\label{fig:QPD}
\end{figure}

Decoupling the dipole interaction in Eq.~\eqref{eq:Dicke} leads to the mean-field Hamiltonian: $\hat{H}_{MF} = \hat{H}_{ph} + \hat{H}_{at}-\frac{2g\alpha m}{\sqrt{N}}$, where $\hat{H}_{ph} = \omega\hat{a}^\dagger\hat{a}+\frac{gm}{\sqrt{N}}(\hat{a}^\dagger+\hat{a})$ and $ \hat{H}_{at} = \epsilon{\hat{S}_{z}} + \frac{2g\alpha}{\sqrt{N}}\hat{S}_x$. This mean-field model is diagonalized by the displacement, $\ahat \rightarrow \ahat - \frac{gm}{\omega\sqrt{N}}$, of the photon operator, and the rotation of the spin operators as in Eq.~\eqref{eq:rotationsfromStoJ}, by the rotation angle, $\theta=\tan^{-1}\left(\frac{2g\alpha}{\epsilon\sqrt{N}}\right)$. The mean-field ground state, $|0\rangle\otimes|J,-J\rangle$, corresponds to having zero photons in the displaced basis and $-J$ eigenvalue for the spin operator $\hat J_z$ in the rotated frame. The two order parameters in the mean-field ground state for $\xi<1$ are:
\begin{subequations}
\label{eq:OP-Dicke}
\begin{eqnarray}	 
m &=& \pm\frac{N}{2}\sqrt{1-\xi^2} \label{eq:m-Dicke} \\
\alpha &=& \frac{-gm}{\omega\sqrt{N}} \label{eq:alpha}
\end{eqnarray} \end{subequations}
where $\xi=\frac{\epsilon\omega}{g^2} $ is a dimensionless parameter. With these non-zero values of $m$ and $\alpha$ for $\xi<1$, the system is in the superradiant phase. The two order parameters vanish continuously at the critical point, $\xi_c=1$. For $\xi \ge 1$, the system is in the normal phase with $m=\alpha=0$. Note that the $m$ in Eq.~\eqref{eq:m-Dicke} is same as that in Eq.~\eqref{eq:OP-m}. In the space of two dimensionless parameters, $\omega/g$ and $\epsilon/g$, the critical line given by $\frac{\epsilon\,\omega}{g^2}=1$ is a hyperbola. See Fig.~\ref{fig:QPD} for the quantum phase diagram of the Dicke model.

\subsection{Bosonic fluctuations in the large $N$ limit}
We now formulate a theory of quantum fluctuations with respect to the mean-field phases of the Dicke model using Holstein-Primakoff transformation for the spin operators. The HP representation has proved to be quite effective in studying Dicke model~\cite{Emary2003}. In the large $N$ limit, it provides a soluble bosonic theory of the Dicke model. We start by applying to the Dicke model in Eq.~(\ref{eq:Dicke}) the same spin rotation and the radiation displacement as in the mean-field theory leading to Eqs.~\eqref{eq:OP-Dicke}. Then, we map the spin operators to bosons using the Holstein-Primakoff transformation: 
$\hat{J}_{z} = -\frac{N}{2} +\bhat^\dag\bhat$ 
and $\hat{J}^+ = \hat{b}^\dagger \sqrt{N-\bhat^\dag\bhat}$. 
In the large $N$ limit, for small spin deviations $\langle \bhat^\dag\bhat \rangle$ $\ll N$, the $\Jhat^+$ is approximated to $\hat J^+ \approx \sqrt{N} \, \bhat^\dag$, and the interaction between the photons and the spin deviations is neglected. By keeping only the bilinear terms in the creation/annihilation operators, we obtain the following bosonic Hamiltonian for the Dicke model:
\begin{eqnarray}
\hat H &=& \omega \, \ahat^\dag\ahat + \varepsilon \, \bhat^\dag\bhat + \gamma \big(\ahat^\dag+\ahat\big)\big(\bhat^\dag+\bhat\big) + e_0
\label{eq:H-boson-Dicke}  
\end{eqnarray}
where $\varepsilon= \epsilon$, $\gamma = g/2$, and $e_0= -\epsilon N/2$ in the normal phase for $\xi\ge 1$, and $\varepsilon= g^2/\omega$, $\gamma = \epsilon\omega/2g$, and $e_0=-\frac{Ng^2}{4\omega}(1+\xi^2)$ in the superradiant phase for $\xi<1$. The constant $e_0$ here is the energy of the mean-field state described in Sec.~\ref{sec:Dicke-MFT}. This Hamiltonian can be diagonalized by the Bogoliubov transformation $\hat U = \Uhat_1\Uhat_2\Uhat_a\Uhat_b$, where $\Uhat_1 = e^{-\phi_1(\ahat^\dag\bhat-\bhat^\dag\ahat)}$, $\Uhat_2 = e^{-\phi_2(\ahat^\dag\bhat^\dag-\bhat \ahat)}$, $\Uhat_a = e^{-\frac{\phi_a}{2}(\ahat^\dag\ahat^\dag-\ahat\ahat)}$ and $\Uhat_b = e^{-\frac{\phi_b}{2}(\bhat^\dag\bhat^\dag-\bhat\bhat)}$, for the Bogoliubov angles given by $\tan{2\phi_1}=\frac{2\gamma}{\omega-\varepsilon}$, $\tanh{2\phi_2}=\frac{2\gamma \cos{2\phi_1}}{\omega+\varepsilon}$, $\tanh{2\phi_a} = \frac{\gamma\sin{2\phi_1}}{\omega_a}$ and  $\tanh{2\phi_b} = -\frac{\gamma\sin{2\phi_1}}{\omega_b}$, where $\omega_{a(b)} = \sqrt{\left(\frac{\omega+\varepsilon}{2}\right)^2-\left(\gamma\cos{2\phi_1}\right)^2} \pm \sqrt{\left(\frac{\omega-\varepsilon}{2}\right)^2+\gamma^2}$. 

The resulting diagonal Hamiltonian, $\mathscr{H}=\hat{U}^\dag\Hhat\hat{U}$, is
\begin{equation}\label{eq:generalDiagHamiltonian}
\mathscr{H} = \Omega_a \left[\ahat^\dag\ahat + \frac{1}{2}\right] +\Omega_b\left[\bhat^\dag\bhat+\frac{1}{2}\right] + e_0-\frac{\omega+\varepsilon}{2}
\end{equation}
where $\Omega_{a(b)}=\sqrt{\omega^2_{a(b)}-(\gamma\sin{2\phi_1})^2}$ are the frequencies (energies) of the `polaronic' (mixed up atom-photon) normal modes, 
whose explicit forms are given below. In the superradiant phase:
\begin{eqnarray}
\Omega_{a(b)} &=& \sqrt{\left(\frac{\omega^4 + g^4}{2\omega^2}\right) \pm \sqrt{\left(\frac{\omega^4 - g^4}{2\omega^2}\right)^2 + \epsilon^2\omega^2}} \label{eq:Omega-ab-super}
\end{eqnarray}		
and in the normal phase:
\begin{eqnarray}
\Omega_{a(b)} &=& \sqrt{\left(\frac{\omega^2 + \epsilon^2}{2}\right) \pm \sqrt{\left(\frac{\omega^2 - \epsilon^2}{2}\right)^2 + g^2 \epsilon \omega}} \label{eq:Omega-ab-normal}
\end{eqnarray}
In the above equations, $\Omega_a$ corresponds to the expressions with `$+$' sign, and $\Omega_b$ to the expressions with `$-$' sign.
\subsection{Dynamics of squeezing across phase transition}
We study the time evolution of squeezing of spin and radiation in the reference state, $|0\rangle_a\otimes |0\rangle_b \equiv |0,0\rangle$, with respect to the Hamiltonian given by Eq.~\eqref{eq:H-boson-Dicke}. This reference state is the mean-field state, $|0\rangle\otimes|J,-J\rangle$, described in Sec.~\ref{sec:Dicke-MFT}, whose quantum fluctuations in the large $N$ limit are described by Eq.~\eqref{eq:H-boson-Dicke}. Following Eq.~\eqref{eq:zetas-HP}, we define below the squeezing parameters $\zeta_s(t)$ and $\zeta_p(t)$ for spin and radiation, respectively. 
\begin{subequations}
\begin{eqnarray} \label{eq:squeezingForm}
\zeta_s(t) &=& \sqrt{\left\langle 0,0\right| e^{i\Hhat t} \left(\bhat^\dag e^{-i\phi_s} + \bhat\,e^{i\phi_s}\right)^2 e^{-i\Hhat t}\left|0,0\right\rangle}_{\rm min} \nonumber \\
&=& \sqrt{C_s(t) -\sqrt{A_s^2(t) + B_s^2(t)}} \label{eq:zetas-Dicke} \\
\zeta_p(t) &=&  \sqrt{\left\langle 0,0\right| e^{i\Hhat t} \left(\ahat^\dag e^{-i\phi_p} + \ahat\,e^{i\phi_p}\right)^2 e^{-i\Hhat t} \left| 0,0 \right\rangle}_{\rm min} \nonumber \\
&=& \sqrt{C_p(t) -\sqrt{A_p^2(t) + B_p^2(t)}} \label{eq:zetap-Dicke}
\end{eqnarray}
\end{subequations}
Here, too, $\left\langle 0,0\right| e^{i\Hhat t} \left(\bhat^\dag e^{-i\phi_s} + \bhat\,e^{i\phi_s}\right) e^{-i\Hhat t}\left|0,0\right\rangle$ as well as $\left\langle 0,0\right| e^{i\Hhat t} \left(\ahat^\dag e^{-i\phi_p} + \ahat\,e^{i\phi_p}\right) e^{-i\Hhat t}\left|0,0\right\rangle$ are zero, and $\phi_{{\rm min}, s(p)}(t) = \frac{\pi}{2} + \frac{1}{2 } \tan^{-1}\left[\frac{B_{s(p)}(t)}{ A_{s(p)}(t)} \right]$. 
For the $\Hhat$ given by Eq.~\eqref{eq:H-boson-Dicke}, the expressions we derived for the parameters $A_{s(p)}$, $B_{s(p)}$ and $C_{s(p)}$ are given below.  
\begin{widetext} 
\begin{subequations}
\label{eq:ABC-spin-Dicke}
\begin{eqnarray}
A_s(t)  &=& \langle 0,0|e^{i\Hhat t} \left(\bhat^\dag\bhat^\dag+\bhat\bhat\right)e^{-i\Hhat t}|0,0 \rangle \\
&=& -\frac{1}{2}\cosh{2\phi_2}\left[ \left(\cosh{2\phi_2}-\cos{2\phi_1}\right)\sinh{4\phi_a} \, \sin^2{\Omega_a t} + \left(\cosh{2\phi_2}+\cos{2\phi_1}\right)\sinh{4\phi_b} \, \sin^2{\Omega_b t}\right] \nonumber  \\
&& -\frac{1}{4}\sin{2\phi_1} \, \sinh{4\phi_2}\left(\cos^2{\Omega_a t} +\cosh{4\phi_a} \, \sin^2{\Omega_a t} + \cos^2{\Omega_b t} +\cosh{4\phi_b} \, \sin^2{\Omega_b t}\right)\nonumber \\
&& - \left[\sinh{2\phi_2} \, \cosh(\phi_a+\phi_b)+\sin{2\phi_1} \, \cosh{2\phi_2} \, \sinh(\phi_a+\phi_b)\right]\sinh(\phi_a+\phi_b) \, \sinh{2\phi_2} \, \cos{(\Omega_a-\Omega_b)t} \nonumber \\
&& + \left[\sinh{2\phi_2} \, \sinh(\phi_a+\phi_b)+\sin{2\phi_1} \, \cosh{2\phi_2} \, \cosh(\phi_a+\phi_b)\right]\cosh(\phi_a+\phi_b) \, \sinh{2\phi_2} \, \cos{(\Omega_a+\Omega_b)t} \nonumber \\ 
&& \nonumber \\
B_s(t) & = & -i \langle 0,0|e^{i\Hhat t} \left(\bhat^\dag\bhat^\dag-\bhat\bhat\right)e^{-i\Hhat t}|0,0\rangle \\ 
&=&\frac{1}{2}\cosh{2\phi_2}\left[\left(1-\cos{2\phi_1} \, \cosh{2\phi_2}\right) \sinh{2\phi_a} \, \sin{2\Omega_a t} + \left(1+\cos{2\phi_1} \, \cosh{2\phi_2}\right) \sinh{2\phi_b} \, \sin{2\Omega_b t}\right]  \nonumber \\
&& + \left[\cos{2\phi_1} \, \sinh{2\phi_2} \, \cosh(\phi_a -\phi_b) + \sin{2\phi_1} \, \sinh(\phi_a-\phi_b)\right]\sinh(\phi_a+\phi_b) \, \sinh{2\phi_2} \, \sin(\Omega_a-\Omega_b)t \nonumber \\
&& + \left[\cos{2\phi_1} \, \sinh{2\phi_2} \, \sinh(\phi_a -\phi_b) + \sin{2\phi_1} \, \cosh(\phi_a-\phi_b)\right] \cosh(\phi_a+\phi_b) \, \sinh{2\phi_2} \, \sin(\Omega_a+\Omega_b)t \nonumber \\
&& \nonumber \\
C_s(t)  &=& 1+ 2 \langle 0,0|e^{i\Hhat t} \, \bhat^\dag\bhat \, e^{-i\Hhat t}|0,0\rangle \\
& = & \frac{1}{2}  \cosh{2\phi_2} \left(\cosh{2\phi_2}-\cos{2\phi_1}\right) \left(\cos^2{\Omega_at} + \cosh{4\phi_a} \, \sin^2{\Omega_a t}\right)  \nonumber \\
&& + \frac{1}{2} \cosh{2\phi_2} \left(\cosh{2\phi_2}+\cos{2\phi_1}\right) \left(\cos^2{\Omega_bt} + \cosh{4\phi_b} \, \sin^2{\Omega_b t}\right)  \nonumber \\
&& + \frac{1}{4} \sin{2\phi_1} \, \sinh{4\phi_2} \left(\sinh{4\phi_a} \, \sin^2{\Omega_at} + \sinh{4\phi_b} \, \sin^2{\Omega_bt}\right) \nonumber \\
&& + \left[\sinh{2\phi_2}\,\sinh(\phi_a+\phi_b) + \sin{2\phi_1} \, \cosh{2\phi_2} \, \cosh(\phi_a+\phi_b)\right]\sinh(\phi_a+\phi_b)\, \sinh{2\phi_2}\, \cos(\Omega_a-\Omega_b)t \nonumber \\
&& - \left[\sinh{2\phi_2}\,\cosh(\phi_a+\phi_b) + \sin{2\phi_1} \, \cosh{2\phi_2} \, \sinh(\phi_a+\phi_b)\right]\cosh(\phi_a+\phi_b)\, \sinh{2\phi_2}\, \cos(\Omega_a+\Omega_b)t \nonumber
\end{eqnarray}
\end{subequations}

\begin{subequations}
\label{eq:ABC-photon-Dicke}
\begin{eqnarray}
A_p(t)  &=& \langle 0,0|e^{i\Hhat t} \left(\ahat^\dag\ahat^\dag+\ahat\ahat\right)e^{-i\Hhat t}|0,0 \rangle \\
&=& -\frac{1}{2}\cosh{2\phi_2}\left[ \left(\cosh{2\phi_2}  + \cos{2\phi_1}\right)\sinh{4\phi_a} \, \sin^2{\Omega_a t} + \left(\cosh{2\phi_2}  - \cos{2\phi_1}\right)\sinh{4\phi_b} \, \sin^2{\Omega_b t}\right] \nonumber  \\
&& + \frac{1}{4}\sin{2\phi_1} \, \sinh{4\phi_2}\left(\cos^2{\Omega_a t} +\cosh{4\phi_a} \, \sin^2{\Omega_a t} + \cos^2{\Omega_b t} +\cosh{4\phi_b} \, \sin^2{\Omega_b t}\right)\nonumber \\
&& - \left[\sinh{2\phi_2} \, \cosh(\phi_a+\phi_b) - \sin{2\phi_1} \, \cosh{2\phi_2} \, \sinh(\phi_a+\phi_b)\right]\sinh(\phi_a+\phi_b) \, \sinh{2\phi_2} \, \cos{(\Omega_a-\Omega_b)t} \nonumber \\
&& + \left[\sinh{2\phi_2} \, \sinh(\phi_a+\phi_b) - \sin{2\phi_1} \, \cosh{2\phi_2} \, \cosh(\phi_a+\phi_b)\right]\cosh(\phi_a+\phi_b) \, \sinh{2\phi_2} \, \cos{(\Omega_a+\Omega_b)t} \nonumber \\ 
&& \nonumber \\
B_p(t) & = & -i \langle 0,0|e^{i\Hhat t} \left(\ahat^\dag\ahat^\dag-\ahat\ahat\right)e^{-i\Hhat t}|0,0\rangle \\ 
&=&\frac{1}{2}\cosh{2\phi_2}\left[\left(1 + \cos{2\phi_1} \, \cosh{2\phi_2}\right) \sinh{2\phi_a} \, \sin{2\Omega_a t} + \left(1 - \cos{2\phi_1} \, \cosh{2\phi_2}\right) \sinh{2\phi_b} \, \sin{2\Omega_b t}\right]  \nonumber \\
&& - \left[\cos{2\phi_1} \, \sinh{2\phi_2} \, \cosh(\phi_a -\phi_b) + \sin{2\phi_1} \, \sinh(\phi_a-\phi_b)\right]\sinh(\phi_a+\phi_b) \, \sinh{2\phi_2} \, \sin(\Omega_a-\Omega_b)t \nonumber \\
&& -  \left[\cos{2\phi_1} \, \sinh{2\phi_2} \, \sinh(\phi_a -\phi_b) + \sin{2\phi_1} \, \cosh(\phi_a-\phi_b)\right] \cosh(\phi_a+\phi_b) \, \sinh{2\phi_2} \, \sin(\Omega_a+\Omega_b)t \nonumber \\
&& \nonumber \\
C_p(t)  &=& 1+ 2 \langle 0,0|e^{i\Hhat t} \, \ahat^\dag\ahat \, e^{-i\Hhat t}|0,0\rangle \\
& = & \frac{1}{2}  \cosh{2\phi_2} \left(\cosh{2\phi_2} + \cos{2\phi_1}\right) \left(\cos^2{\Omega_at} + \cosh{4\phi_a} \, \sin^2{\Omega_a t}\right)  \nonumber \\
&& + \frac{1}{2} \cosh{2\phi_2} \left(\cosh{2\phi_2} - \cos{2\phi_1}\right) \left(\cos^2{\Omega_bt} + \cosh{4\phi_b} \, \sin^2{\Omega_b t}\right)  \nonumber \\
&& - \frac{1}{4} \sin{2\phi_1} \, \sinh{4\phi_2} \left(\sinh{4\phi_a} \, \sin^2{\Omega_at} + \sinh{4\phi_b} \, \sin^2{\Omega_bt}\right) \nonumber \\
&& + \left[\sinh{2\phi_2}\,\sinh(\phi_a+\phi_b) - \sin{2\phi_1} \, \cosh{2\phi_2} \, \cosh(\phi_a+\phi_b)\right]\sinh(\phi_a+\phi_b)\, \sinh{2\phi_2}\, \cos(\Omega_a-\Omega_b)t \nonumber \\
&& - \left[\sinh{2\phi_2}\,\cosh(\phi_a+\phi_b) - \sin{2\phi_1} \, \cosh{2\phi_2} \, \sinh(\phi_a+\phi_b)\right]\cosh(\phi_a+\phi_b)\, \sinh{2\phi_2}\, \cos(\Omega_a+\Omega_b)t \nonumber
\end{eqnarray}
\end{subequations}
\end{widetext}
The $A_p$, $B_p$, $C_p$ in Eqs~\eqref{eq:ABC-photon-Dicke} can be obtained from the $A_s$, $B_s$, $C_s$ Eqs.~\eqref{eq:ABC-spin-Dicke} by exchanging $\phi_a$ with $\phi_b$, $\Omega_a$ with $\Omega_b$ and changing $\phi_1$ to $-\phi_1$, while keeping $\phi_2$ unchanged. 

We apply this prescription to calculate squeezing in the two phases of the Dicke model. Our findings from this calculation are presented and discussed below.

\subsection{Results and discussion}
From what we learnt in Sec.~\ref{sec:squeezingOAT-2}, we expect the enhancement of strength and  life of squeezing near the quantum phase boundary to be determined mainly by the smaller of the two normal mode frequencies, $\Omega_{a}$ and $\Omega_b$, given by Eqs.~\eqref{eq:Omega-ab-super} and~\eqref{eq:Omega-ab-normal}. We see that $\Omega_b$ is always smaller of the two, and smoothly tends to $0$ as one approaches the phase boundary from either side. To understand the behaviour of the soft-mode frequency $\Omega_b$ in the close neighbourhood of the critical line, we rewrite $\epsilon/g$ and $\omega/g$ conveniently as: $\omega/g = \sqrt{\xi}\, e^{\psi/2}$ and $\epsilon/g=\sqrt{\xi}\, e^{-\psi/2}$, where $\psi$ is a measure of atom-photon detuning. That is, $\psi=-2\tanh^{-1}{\Delta}$, where $\Delta = (\epsilon-\omega)/(\epsilon+\omega)$ is relative detuning. For instance, $\psi\rightarrow +\infty$ (i.e., $\Delta\rightarrow -1$) corresponds to having $\omega \gg \epsilon$, and $\psi\rightarrow - \infty$ (i.e., $\Delta\rightarrow +1$) gives the opposite extreme, $\omega\ll \epsilon$. The resonant case, $\omega=\epsilon$, is given by $\psi=\Delta=0$. Now, if $\delta=|\xi-1|$ is a small distance from the phase boundary, then to the leading order in $\delta$, we find that $\Omega_b \approx g\sqrt{\delta/2\cosh{\psi}}$ in the normal phase, and $\Omega_b \approx g \sqrt{\delta/\cosh{\psi}}$ in superradiant phase. Hence, the time period of squeezing oscillations grows as $|\xi-\xi_c|^{-1/2}$. This power-law behaviour for the critical growth of the squeezing time for the Dicke model is same as in Eq.~\eqref{eq:powerlawOAT} for the OAT model in transverse field, but with a factor that depends on the detuing angle $\psi$ which suggests that the critical enhancement of squeezing time will be better achieved in the highly detuned cases. 

\begin{figure}[h!]
\centering
\includegraphics[width=.5\textwidth]{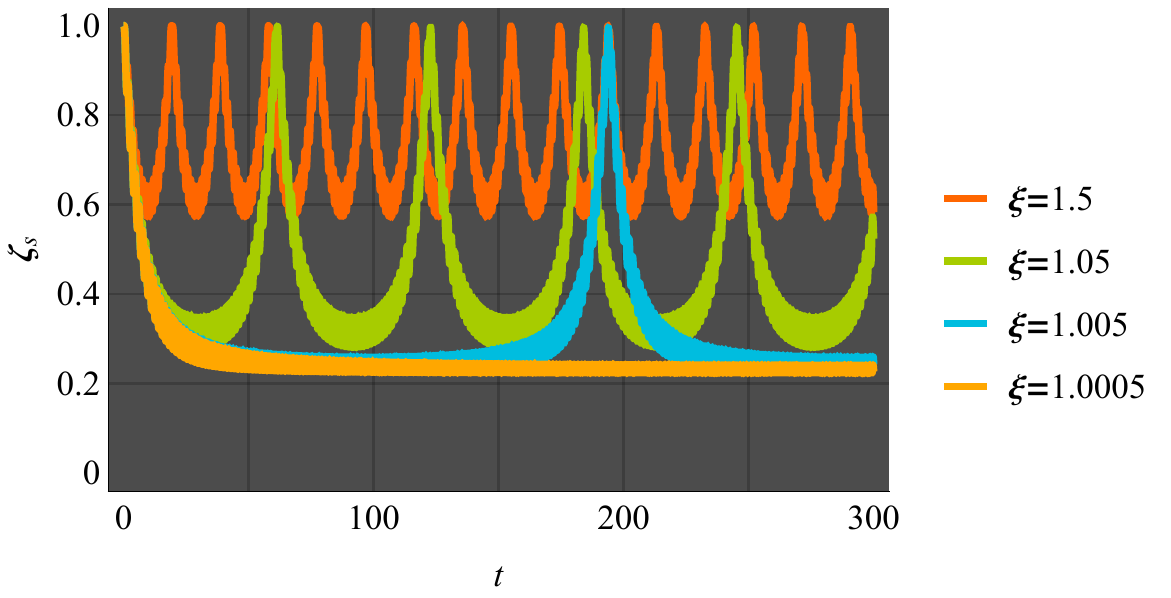} \\
\vspace{4mm}
\includegraphics[width=.5\textwidth]{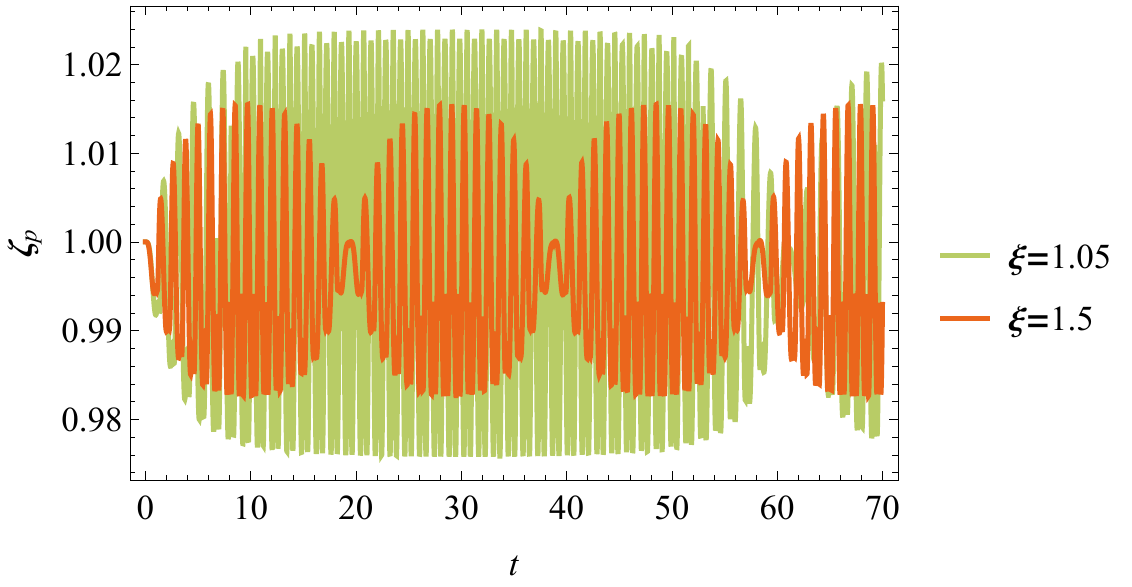}
\caption{Squeezing parameters $\zeta_s$ (for spin) and $\zeta_p$ (for photon) in the normal phase ($\xi>1$) of the Dicke model for spin-like detuning ($\Delta=-0.9$). The $\zeta_s$ exhibits a sharp reduction in its minimum value (i.e., high degree of spin squeezing) and a great elongation in the squeezing time, as $\xi$ approaches $1$ (the critical value). The $\zeta_p$ stays close to 1 while undergoing rapid oscillations and beating (i.e., no photon squeezing).}
\label{fig:zetasp-Normal-spinlike}
\end{figure}

\begin{figure}[h!]
\centering
\includegraphics[width=.5\textwidth]{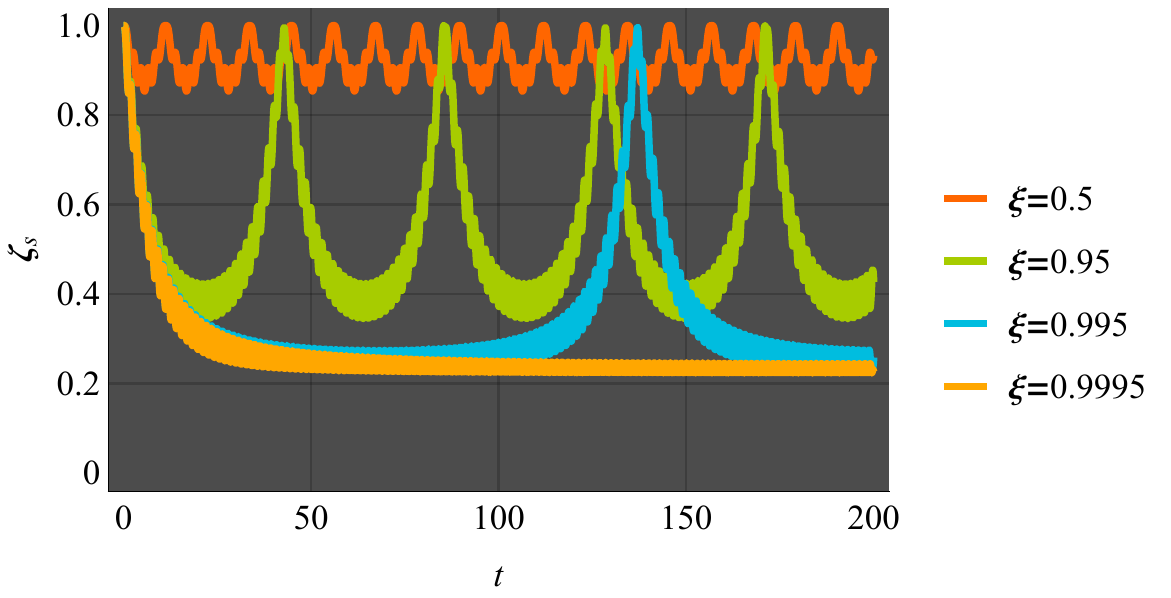} \\
\vspace{4mm}
\includegraphics[width=.5\textwidth]{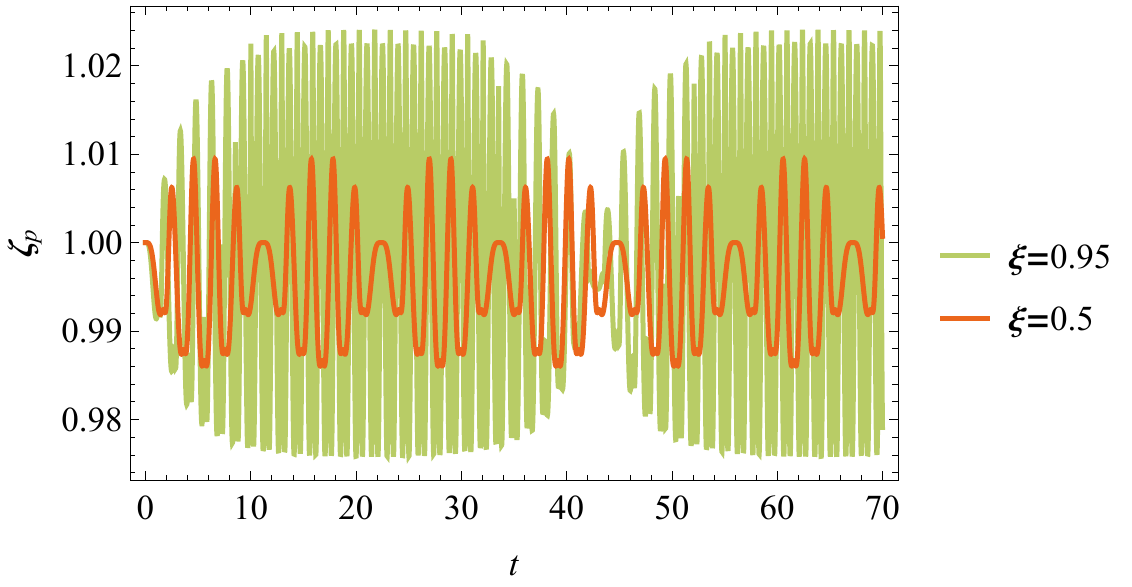}
\caption{Squeezing parameters for spin ($\zeta_s$) and photon ($\zeta_p$) in the superradiant phase ($\xi<1$) of the Dicke model for spin-like detuning ($\Delta=-0.9$). In this case, the spin again shows a conspicuous enhancement in squeezing time and strength close to the critical value, while the radiation doesn't.}
\label{fig:zetasp-Super-spinlike}
\end{figure}

\begin{figure}[h!]
\centering
\includegraphics[width=.5\textwidth]{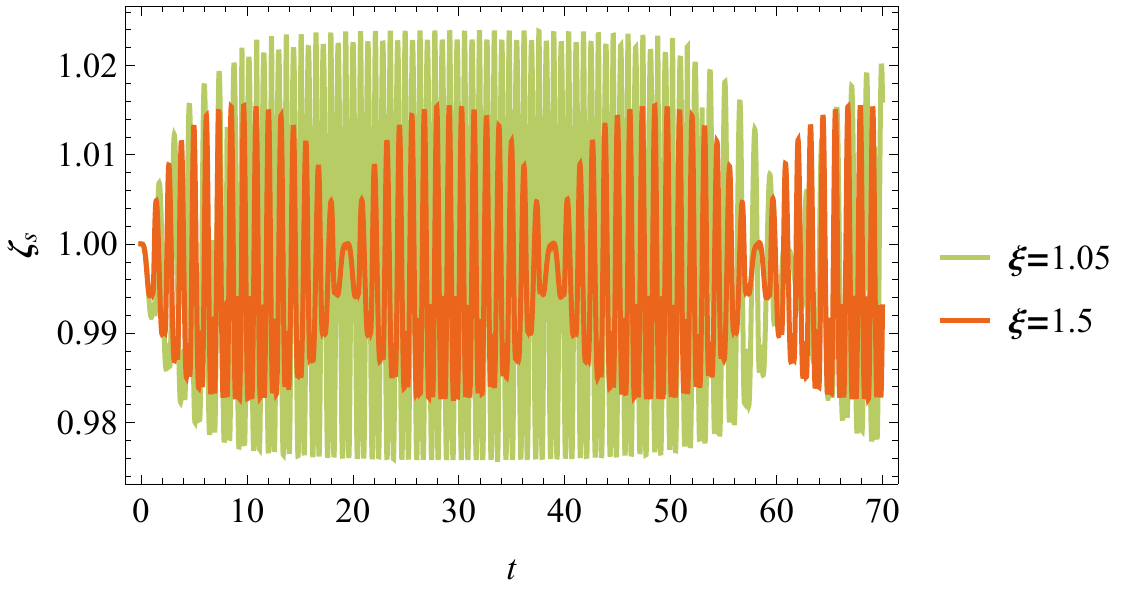} \\
\vspace{4mm}
\includegraphics[width=.5\textwidth]{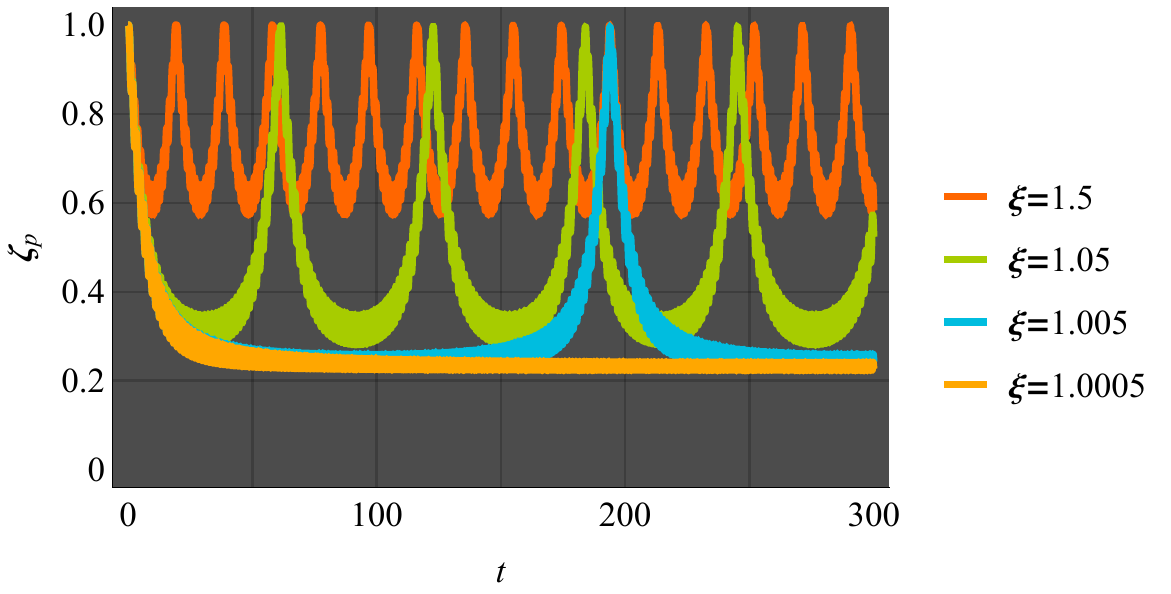}
\caption{Squeezing parameters $\zeta_s$ and $\zeta_p$ in the normal phase of the Dicke model for photon-like detuning ($\Delta=+0.9$). In this case, the exact opposite of what we saw in Fig.~\ref{fig:zetasp-Normal-spinlike} happens. Here, the radiation shows conspicuous squeezing effects near the critical point, but not the spin.}
\label{fig:zetasp-Normal-photonlike}
\end{figure}

\begin{figure}[h!]
\centering
\includegraphics[width=.5\textwidth]{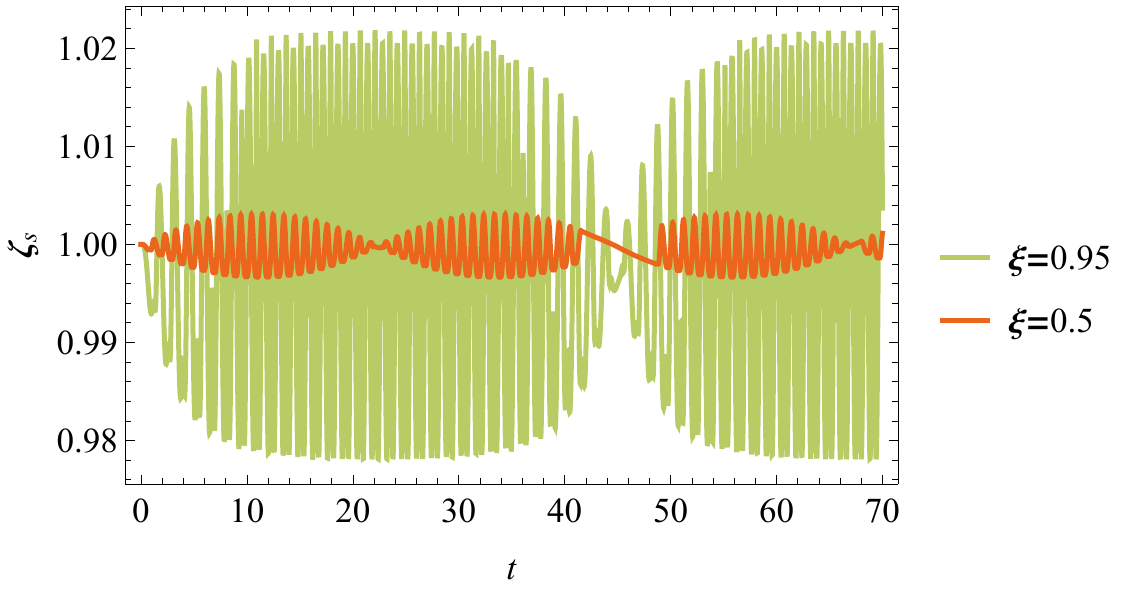} \\
\vspace{4mm}
\includegraphics[width=.5\textwidth]{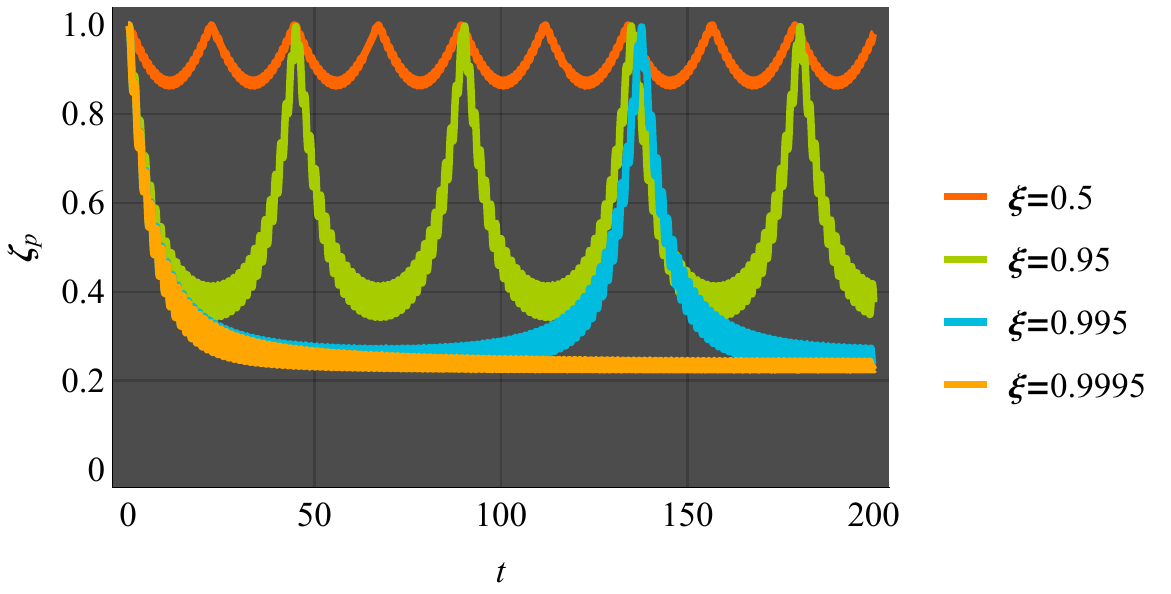}
\caption{Squeezing parameters $\zeta_s$ and $\zeta_p$ in the superradiant phase of the Dicke model for photon-like detuning ($\Delta=+0.9$). Here, the photons shows a strong enhancement in the squeezing strength and time near the critical point.}
\label{fig:zetasp-Super-photonlike}
\end{figure}

Note that, along a fixed $\xi$ contour in the normal phase, $\Omega_b \approx \epsilon \sqrt{1-\xi^{-1}}$ for $\epsilon \ll \omega $. That is, $\Omega_b$ in this highly detuned limit is spin-like~\footnote{Deep inside the normal phase (i.e., for sufficiently large $\xi$), $\Omega_b \approx \epsilon$ for $\omega \gg \epsilon$. Hence, we call this normal mode spin-like. In the same way, $\Omega_b \approx \omega$ for $\epsilon \gg \omega$ is termed photon-like.}. Likewise, $\Omega_b \approx \omega \sqrt{1-\xi^{-1}}$ for $\omega \ll \epsilon$ is photon-like. A similar consideration in the superradiant phase is a bit tricky, because $\varepsilon$ for the spin fluctuations in Eq.~\eqref{eq:H-boson-Dicke} is not equal to $\epsilon$, but $g^2/\omega$. Keeping this in mind, for $\epsilon \ll \omega$ along a fixed $\xi$ contour in the superradiant phase, $\Omega_b \approx \frac{g^2}{\omega} \sqrt{1-\xi^2}$  is indeed spin-like, and $\Omega_b \approx \omega\sqrt{1-\xi^2} $ for $\omega \ll \epsilon$ is photon-like. Thus, close to the critical line, spin squeezing is expected to grow stronger and live longer for $\epsilon$ much smaller than $\omega$ (i.e., when the relative detuning $\Delta$ is closer to $-1$). On the other hand, for $\omega$ much smaller than $\epsilon$ (i.e., $\Delta$ closer to $+1$), the photon squeezing will get enhanced and be sustained for longer times. The data in Figs.~\ref{fig:zetasp-Normal-spinlike} to \ref{fig:zetasp-Super-photonlike} for $\zeta_s$ and $\zeta_p$ given by Eqs.~\eqref{eq:zetas-Dicke} and~\eqref{eq:zetap-Dicke} [and using Eqs.~\eqref{eq:ABC-spin-Dicke} and~\eqref{eq:ABC-photon-Dicke}] shows this to be true. 

\subsubsection{Spin squeezing for spin-like detuning}
Figures~\ref{fig:zetasp-Normal-spinlike} and~\ref{fig:zetasp-Super-spinlike} present $\zeta_s(t)$ and $\zeta_p(t)$ in the normal and the superradiant phase respectively, for the spin-like detuning of $\Delta=-0.9$ for different values of $\xi$. Here, we see that the spin squeezing parameter $\zeta_s$ is always less than 1, and shows high reduction in its minimum value as $\xi$ approaches the phase boundary from either side. Closer to the critical point, we also see the striking appearance of flatness in $\zeta_s$ around its minimum value (say, $\zeta_{s,\rm min}$) for the most part of time in every squeezing cycle with greatly increased time period (due to the softening of $\Omega_b$). Thus, across the superradiant transition in the Dicke model with spin-like detuning, we see a marked growth in the time and strength of spin squeezing near the critical point. These findings for the Dicke model are very much like what we got for the OAT model in the previous section. But there are a few notable differences too, as described below. 

For a given detuning $\Delta$ (or $\psi$), the $\zeta_{s,\rm min}$ doesn't vanish as $\xi$ tends to 1. Instead, it saturates to a non-zero value. See the minimum values of $\zeta_s$ for $|\xi-1|=0.05, 0.005$ and $0.0005$ in Figs.~\ref{fig:zetasp-Normal-spinlike} and~\ref{fig:zetasp-Super-spinlike}. But we find the $\zeta_{s,\rm min}$ to decrease with increasing $\Delta$ (see Fig.~\ref{fig:CriticalLaw-NS-spin} and the related discussion for more on this). Another notable point is that the flat looking parts of $\zeta_s(t)$ are not quite flat. They are modulated by weak high frequency oscillations (due to $\Omega_a$). The amplitude of these rapid oscillations is found to decrease sharply as $\xi$ approaches 1. So, in the close proximity of the critical point, we essentially have a constant small value of $\zeta_s$ over a long duration of time, i.e. strong spin squeezing with a long life. 

The photon squeezing parameter $\zeta_p$ in Figs.~\ref{fig:zetasp-Normal-spinlike} and~\ref{fig:zetasp-Super-spinlike} oscillates extremely rapidly (and beats) around 1 with a very small amplitude. Hence, practically there is no photon squeezing in the Dicke model with spin-like detuning. 

Although not shown here, but the prominent squeezing effects described above for spin are found to weaken and disappear as $\Delta$ is reduced and taken closer to 0, i.e. near resonance. Thus, the high spin-like detuning favours spin squeezing, and exhibits critical enhancements. But the situation is found to reverse, as described below, when we consider photon-like detuning, i.e. $\Delta$ close to $+1$. 

\subsubsection{Photon squeezing for photon-like detuning}
Figure~\ref{fig:zetasp-Normal-photonlike} and~\ref{fig:zetasp-Super-photonlike} present the squeezing parameters for the photon-like detuning of $\Delta=+0.9$ in the two phases. Here, on approaching the critical point in either phase, not spin but photon squeezing parameter $\zeta_p$ shows a marked reduction in its minimum value, and a great increase in the time period of oscillation, with a largely flat $\zeta_p$ (around the minimum value) in every squeezing cycle. All that we learnt about the critical enhancement of spin squeezing in the case of spin-like detuning also holds true for the critical enhancement of photon squeezing in the case of photon-like detuning. In fact, in the normal phase, the $\zeta_p$ for photon-like detuning exactly replicates the $\zeta_s$ for spin-like detuning. Compare $\zeta_s$ ($\zeta_p$) in Fig.~\ref{fig:zetasp-Normal-spinlike} with $\zeta_p$ ($\zeta_s$) in Fig.~\ref{fig:zetasp-Normal-photonlike}. They are exactly the same because, in the normal phase, there exists a symmetry between spin and photon under $\Delta \rightarrow -\Delta$. Even in the superradiant phase, where there is no exact symmetry relating the photon with the spin for opposite detuning, close to the critical line, $\zeta_p$ for positive $\Delta$ still behaves pretty much like the $\zeta_s$ for spin-like detuning. Compare, for instance, the $\zeta_s$ in Fig.~\ref{fig:zetasp-Super-spinlike} with $\zeta_p$ in Fig.~\ref{fig:zetasp-Super-photonlike}. They are nearly the same, with same qualitative features. Hence, for photon-like detuning, it is the radiation which exhibits prominent enhancements in squeezing time and strength, while the atoms (spin) show no such effects. 

\subsubsection{Critical behaviour of squeezing}
The critical behaviour of the growth of squeezing time for spin and photon (with appropriate detuning) is governed by the frequency, $\Omega_b$, of the soft polariton mode. We have shown earlier that, close to the critical point, $\Omega_b \sim |\xi-\xi_c|^{1/2}$. Therefore, the time for which the system stays in the enhanced squeezed state (i.e., the flat part of the squeezing cycle $\propto \Omega_b^{-1}$) grows as $|\xi-\xi_c|^{-1/2}$. This is exactly like the critical behaviour of the squeezing time for the OAT model in transverse field. See Eq.~\eqref{eq:T-powerlawOAT}. 

\begin{figure}[h!]
\centering
\includegraphics[width=.5\textwidth]{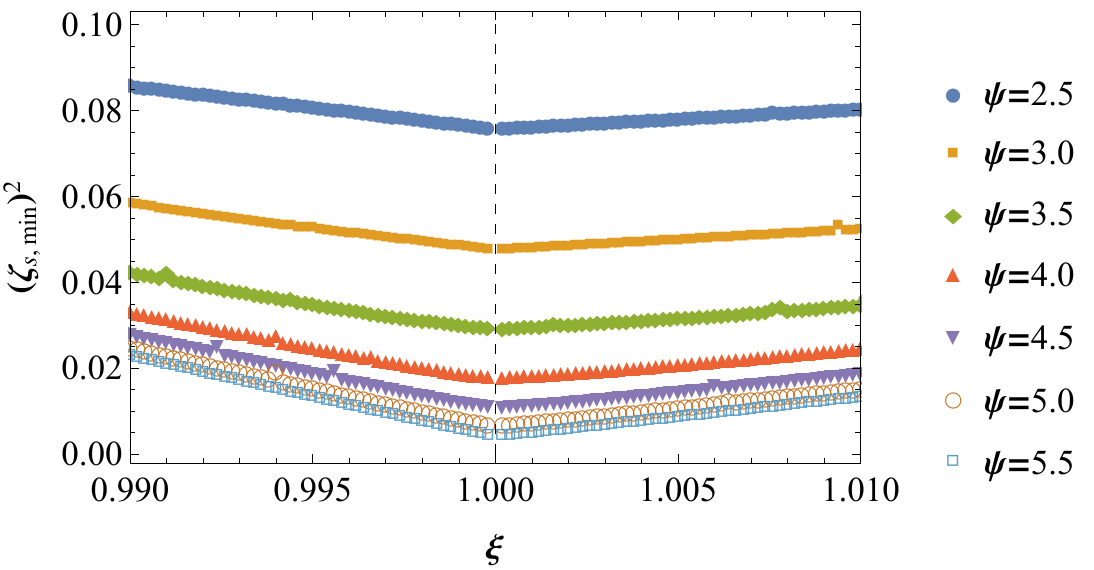}
\caption{Critical behaviour of spin squeezing across the superradiant transition in the Dicke model. This data suggests that $(\zeta_{s,{\rm min}})^2 = u + v\, |\xi-1| $ in the close proximity of the critical point, where the intercept $u$ tends to $0$ with increasing $\psi$ (i.e., $\Delta\rightarrow -1$) and the slope $v$ in the two phases is different.}
\label{fig:CriticalLaw-NS-spin}
\end{figure}

To understand the critical behaviour of the growth of squeezing strength, let us carefully look at $\zeta_s$ for spin-like detuning near the critical point. Whatever we learn from this would also apply to $\zeta_p$ for photonlike detuning. In Fig.~\ref{fig:CriticalLaw-NS-spin}, we have plotted the square of the minimum value of $\zeta_s$, i.e $(\zeta_{s,\rm min})^2$, as a function of $\xi$ in the close neighbourhood of the critical point ($\xi_c=1$, marked by the dashed line) for different values of detuning (given here by $\psi$; recall that $\Delta=-\tanh{\frac{\psi}{2}}$). From this figure, it is clear that  
$(\zeta_{s,\rm min})^2 = u +v\, |\xi-\xi_c|$, near the critical point. It is also clear that the slope $v$ in the superradiant phase is bigger than that in the normal phase. Moreover, $u\rightarrow 0$, as $\Delta\rightarrow -1$. Hence, the critical behaviour of the strength of spin squeezing in the Dicke model with spin-like detuning is given by 
\begin{equation}
\zeta_{s,\rm min}=\sqrt{u+v\, |\xi-\xi_c|}.
\label{eq:criticallaw}
\end{equation}
This is different from the $|\xi-\xi_c|^{1/2}$ behaviour obtained for the OAT model in transverse field. For extremely large detuning (i.e. $u \rightarrow 0 $), it does becomes $|\xi-\xi_c|^{1/2}$, which is consistent with the OAT model. See Eq.~\eqref{eq:zetasmin-powerlawOAT}. But in general for a non-zero $u$ (i.e., for not-so-extremely detuned cases), the $\zeta_{s,\rm min}$ infinitesimally close to the critical point behaves as $|\xi-\xi_c|$ plus a constant, showing a change of critical exponent from 1/2 to 1. The same would also apply to the critical behaviour of photon squeezing in the Dicke model with photon-like detuning.
 
\section{Conclusion}\label{sec:sum}
In this paper, we have studied the dynamics of squeezing across quantum phase transition in the OAT model in transverse field and the Dicke model. These models undergo continuous transition from the disordered (normal) to ordered (superradiant) phase with change in a suitable model parameter $\xi$. We have used Holstein-Primakoff transformation in the large spin limit to formulate an exactly doable bosonic theory, and deduced the power-law behaviour for the enhancement of squeezing time and strength near the quantum critical point $\xi_c$. We have shown the amplitude of squeezing parameter to behave as $|\xi-\xi_c|^{1/2}$ for the OAT model, and as $\sqrt{u + v\, |\xi-\xi_c|}$ for the Dicke model, which for extremely large detuning also becomes $|\xi-\xi_c|^{1/2}$. The squeezing time in both the models is found to grow as $|\xi-\xi_c|^{-1/2}$. In the Dicke model, these strong squeezing effects for spin or photon are seen respectively in the spin-like or photon-like detuned cases. To conclude, we have shown that it is possible to attain high degree of squeezing for a long duration of time near quantum critical points, and presented an understanding of its critical behaviour. It is desirable to further look at the effect of dissipation, temperature and finiteness of spin on these findings, and in different models.
\begin{acknowledgments}
D.S. acknowledges the financial support from Jawaharlal Nehru University (JNU) during her PhD. B.K. acknowledges general financial support under the DST (India) funded PURSE program of JNU.  
\end{acknowledgments}	

\bibliography{references}

\end{document}